\input{aipcheck}

\documentclass[
    ,final            % use final for the camera ready runs
  ]
  {aipproc}

\layoutstyle{8x11double}

\usepackage{amssymb}
\usepackage{amsmath}

\newcommand{\order}{\mathcal{O}}

\newcommand{\GeV}{\ \mathrm{GeV}}

\newcommand{\Mgrav}{M_*}
\newcommand{\MGUT}{M_\mathrm{GUT}}

\newcommand{\tb}{\tan\!\beta}
\newcommand{\cb}{\cos\!\beta}
\newcommand{\tmg}{\tau \rightarrow \mu \gamma}

\newcommand{\meg}{\mu \rightarrow e \gamma}
\newcommand{\bsg}{B \rightarrow X_s \gamma}

\newcommand{\bsbsbar}{$B_s$--$\overline{B_s}$}

\newcommand{\SphiK}{S_{CP}^{\phi K}}

\newcommand{\deltad}{\delta^d}

\newcommand{\ded}[2]{(\deltad_{#1})_{#2}}

\newcommand{\incgr}[2][]{\includegraphics[#1]{#2.pdf}}

\newcommand{\arXivid}[1]{arXiv:#1}

\begin{document}

\title{\boldmath$B_s$
 mixing phase and lepton flavor violation in supersymmetric SU(5)}


\author{Jae-hyeon Park}{
  address={INFN, Sezione di Padova, via F Marzolo 8, I--35131, Padova, Italy}
}

% \author{<author2>}{
%   address={<common address for author2 and author3>}
% }
% 
% \author{<author3>}{
%   address={<common address for author2 and author3>}
%   ,altaddress={<author1 address>} % additional visiting address
% }

\begin{abstract}
  The connection between $B_s$ mixing phase
  and lepton flavor violation is studied in SU(5) GUT\@.
  The $\order(1)$ phase, preferring a non-vanishing squark mixing,
  generically implies $\tau \rightarrow (e + \mu)\,\gamma$ and $\meg$.
  In addition to the facts already well-known,
  stresses are put on
  the role of gaugino to scalar mass ratio at the GUT scale and
  the possible modifications
  due to Planck-suppressed non-renormalizable operators.
\end{abstract}

\maketitle

%%%%%%%%%%%%%%%%%%%%%%%%%%%%%%%%%%%%%%%%%%%%
%% MAINMATTER
%%%%%%%%%%%%%%%%%%%%%%%%%%%%%%%%%%%%%%%%%%%%

The $B_s$ mixing phase, denoted by $\phi_s$, is a theoretically clean
observable, and one can make a close
connection between its data and a theory possibly involving new physics.
% We choose the notation of $\phi_s$ used by the D\O\ collaboration.
In the SM, one has $\phi_s \simeq - 2 \eta \lambda^2 \simeq -0.04$.
On the experimental side,
the latest world average of the data
from D\O\ \cite{Abazov:2008fj} and CDF \cite{Aaltonen:2007he},
reported by the UTfit collaboration \cite{Bona:2008jn},
appears to favor a negative $\order(1)$ value of $\phi_s$,
although the measurement is still at an early stage.
% This result shows a discrepancy of $3.7\sigma$
% between the data and the SM prediction.
If the difference solidifies, it should be a clean
indication of a new source of $CP$ violation.

% SUSY has a lot of sources
%% scalar masses, A terms
A supersymmetric extension of the SM has potential new sources of
flavor and/or $CP$ violation in its soft supersymmetry breaking terms.
It might be conceivable that one of them is revealing its existence
through the above anomaly.
We employ the notation of mass insertion parameters,
written in the form of $\ded{ij}{AB}$ with
the generation indices $i, j = 1,2,3$ and the chiralities $A,B = L,R$.
We do not only use their usual definition at the weak scale,
but also borrow the same notation to specify an off-diagonal element
of the soft scalar mass matrix at $\MGUT$,
the unification scale \cite{fcncgut}.
Being a transition between the second and the third families,
\bsbsbar\ mixing is naturally associated with $\ded{23}{AB}$.
%% LL and/or RR could give large contributions to mixing
Among the four possibilities,
the $LR$ and the $RL$ mass insertions tend to cause
an unacceptable change in $\bsg$
before they can give an appreciable modification
to \bsbsbar\ mixing \cite{LRRLbsgam}.
Therefore, we focus on $LL$ and $RR$ mixings in what follows.

% intro of GUT
In this work, we
work with a grand unified theory (GUT)\@.
%% relation between squark mixing and slepton mixing
Since a single GUT multiplet contains both (s)quarks and (s)leptons,
flavor transitions in the two sectors are related.
%% UTfit implies LFV
Then, one immediately arrives at the
conclusion that the new source of $b \leftrightarrow s$ transition,
needed to account for $\phi_s$,
generically implies lepton flavor violation (LFV)\@.
%% we look at this
We wish to consider this scenario in a model independent fashion
taking SU(5) as the unified gauge group.

If one has a perfect alignment between the mass eigenstates of
quarks and leptons,
$\ded{ij}{RR}$ implies the transition of $l_j \rightarrow l_i$.
However, this straightforward correspondence may be broken by
the inclusion of non-renormalizable terms into the superpotential
as a solution to the wrong quark--lepton mass relations of the lighter two
families.
With the assumption that the cutoff scale of the GUT is
two orders of magnitude higher than $\MGUT$,
one can nevertheless constrain $\ded{23}{RR}$
using the combined mode, $\tau \rightarrow (e + \mu)\,\gamma$,
exploiting the fact that the breakdown of $b$--$\tau$ alignment
is suppressed by $\cb$ \cite{fcncgut}.

Another point to note is that $\ded{23}{RR}$ not only leads to $\tmg$
but also $\meg$ thanks to the radiative correction from the top Yukawa
coupling to the scalar mass terms of $\mathbf{10}$
\cite{Hisano:1995cp}.  The $\meg$ constraint
is particularly important for low $m_0$ \cite{fcncgut}.

\begin{figure}[b]
  \centering
  \includegraphics{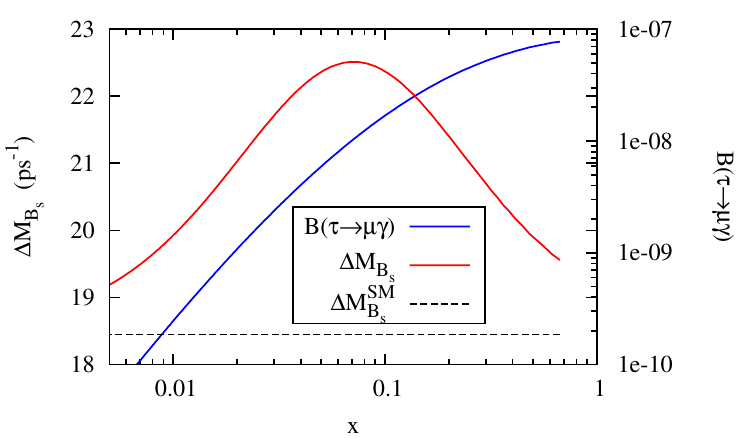}
  \caption{LFV and $B_s$ mixing as functions of $x \equiv M_{1/2}^2 / m_0^2$,
    for fixed $M_{1/2} = 180\GeV$.}
  \label{fig:x}
\end{figure}%
When one uses a low energy hadronic process
to constrain $\delta$ parameters at $\MGUT$,
the running effect of soft mass terms results in an interesting behavior,
illustrated in Fig.~\ref{fig:x}.
% It is due to the competition between squark decoupling and the growth of
% a $\Delta$ parameter,
% as the diagonal components of the squark mass matrix increase.
Suppose that the $\delta$'s and
the gaugino mass $M_{1/2}$ are fixed at $\MGUT$.
Imagine that one increases $m_0$,
the common diagonal entries of soft squark mass matrix at $\MGUT$.
As $m_0$ increases, $(\Delta^d_{ij})_{AB} \equiv m_0^2 \times \ded{ij}{AB}$
grows as well, thereby exerting more and more influence
on low energy flavor violation such as $B_s$ mixing.
At some point, however, squark loop effects begin to decouple
as the squarks become too heavy.
For \bsbsbar\ mixing, this is around $x \equiv M_{1/2}^2 / m_0^2 = 1/12$.
This gaugino to scalar mass ratio could be regarded as a condition
for optimizing the sensitivity of a hadronic process to
flavor non-universality at $\MGUT$ \cite{fcncgut}.
The importance of this observation is more pronounced when one tries
to compare hadronic and leptonic constraints since the latter
is monotonically weakened as $m_0$ is being raised.

\begin{figure}
  \centering
  \incgr[height=60mm]{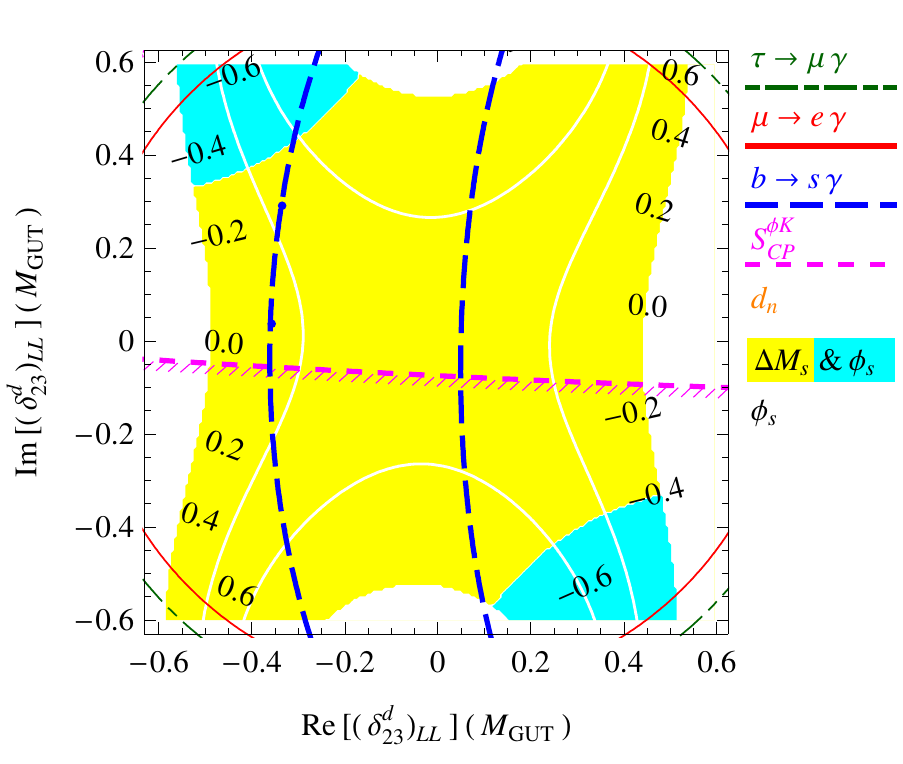}
  \caption{Constraints on $\ded{23}{LL}$.
    For $\tmg$, the thin circle is an upper bound
    from the prospective branching ratio limit, $10^{-8}$.
    For $\meg$, the thin circle shows the projected bound
    on the branching ratio, $10^{-13}$.
    A yellow region is allowed by $\Delta M_s$,
    given 30\% uncertainty in the $\Delta B = 2$ matrix element,
    and a cyan region is further consistent with $\phi_s$.
    Of the two sides of the $\SphiK$ curve,
    the excluded one is indicated by thin short lines.}
  \label{fig:LL}
\end{figure}%
% large extra contribution to Bs demands LL and/or RR
We restrict ourselves to $LL$ and $RR$ mixings of down-type squarks.
%% three cases: LL, RR, LL=RR
We consider three scenarios:
the $LL$ scenario, the $RR$ scenario, and the $LL=RR$ scenario.
%%% scenarios are named after MI's given at MGUT: even in RR scenario,
%%% there is LL from RGE
%%% also RG-induced LL is considered at MGUT
It should be noted that
we set an $LL$ insertion, unless it is a scanning variable,
to a value generated by RG running
from the supersymmetry breaking mediation scale $\Mgrav$
down to $\MGUT$, where
$\Mgrav$ is taken to be the reduced Planck scale.
% is the scale at which supersymmetry breaking is mediated.
% We do this for $\ded{12}{LL}$ and $\ded{13}{LL}$ as well.
These boundary conditions are given at $\MGUT$
with which we solve one-loop RG equations down to the weak scale.
% We consider only the gluino loop contributions to a quark sector process.
% inputs
The constraint from each observable is depicted on
the complex plane of a GUT scale mass insertion.
As for $\phi_s$, we use the 95\% probability region \cite{Bona:2008jn},
\begin{equation}
  \phi_s \in [-1.10, -0.36] \cup [-2.77, -2.07] .
\end{equation}
For concreteness, we assume that there is an
exact quark--lepton flavor alignment.
% In the $RR$ and $LL=RR$ scenarios,
Regarding $\tmg$,
it is straightforward to translate their bounds presented below
to a case with quark--lepton misalignment discussed above---%
interpret $B (\tmg)$ as $B (\tau \rightarrow (e + \mu)\,\gamma)$.
This prescription is applicable to all the three scenarios considered here.
As for $\meg$, barring accidental cancellations,
a contour does not need a modification
in the $RR$ and $LL=RR$ scenarios, while
we do not have a systematic way to account for a misalignment
in the $LL$ scenario.
We fix $M_{1/2} = 180\GeV$, which makes the gluino mass be $500\GeV$
at the weak scale, and set
$m_0 = 600\GeV$, which optimizes the sensitivity of neutral meson mixing
to $\delta$'s at the GUT scale.
We use $\tb = 5$.
Other choices of parameters
are considered in Refs.~\cite{fcncgut},
which also explain other details.

%%%%%%%%%%
% Result %
%%%%%%%%%%

% LL
First, the $LL$ mixing scenario is shown in Fig.~\ref{fig:LL}.
One can find cyan regions that
lead to $\phi_s$ within its 95\% CL intervals.
They involve an $\order(1)$ size of $\ded{23}{LL}$.
However, the supersymmetric disturbance is great also in $\bsg$,
%% low tanb
%% high tanb
%%% worsens bsg SphiK more dangerous
which excludes the bulk of a cyan zone.
The disturbance in this decay mode grows with $\tb$ \cite{Ko:2008xb},
as does that in $\SphiK$.
Although not shown here, $\SphiK$ conflicts with $\phi_s$ for
$\tb = 10$ \cite{fcncgut,Ko:2008xb}.
In this scenario,
discovery of LFV seems to be difficult at a super $B$ factory or MEG\@.

\begin{figure}
  \centering
  \incgr[height=60mm]{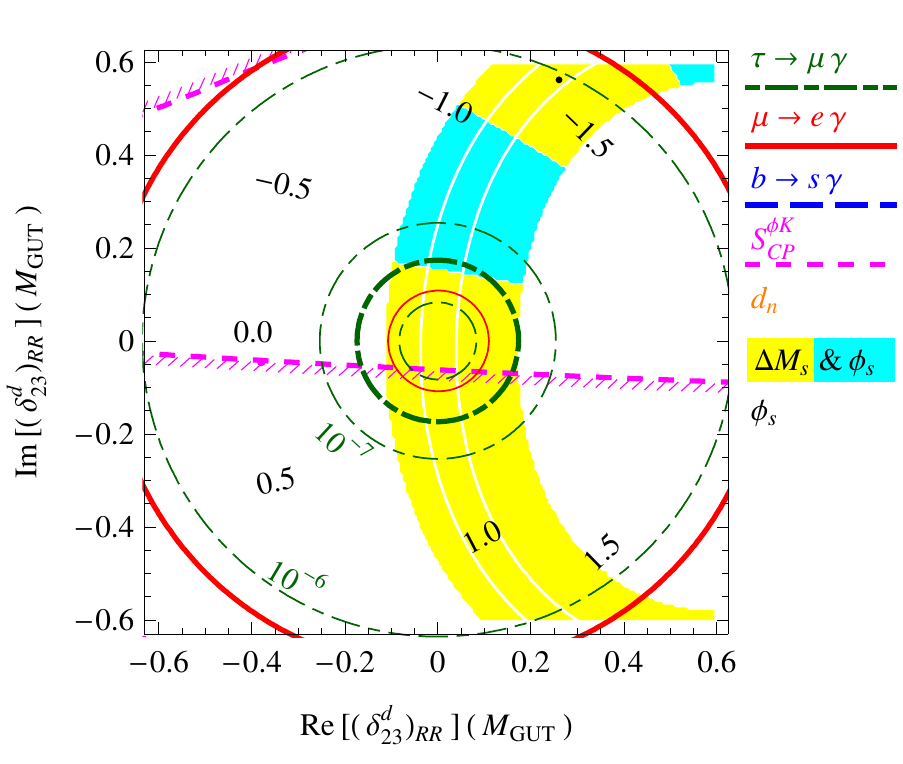}
  \caption{Constraints on $\ded{23}{RR}$.
    For $\tmg$, the thick circle is the current upper bound,
    and the thin circles are, from inside, branching ratios of
    $10^{-8}$, $10^{-7}$, $10^{-6}$, respectively.
    For $\meg$, the thick circle is the current upper bound,
    and the thin circle is branching ratio of $10^{-13}$.
    The remaining marks have the same meanings as in Fig.~\ref{fig:LL}.}
  \label{fig:RR}
\end{figure}

% \rem{elaborate on quark--lepton flavor misalignment}
% \rem{LFV--phase correlation can be read off the contours}
% RR
The $RR$ scenario is shown in Fig.~\ref{fig:RR}.
%% low m0
Comparing this figure with
Fig.~\ref{fig:LL}, one notices that
an $RR$ insertion gives more effect on $B_s$ mixing
than an $LL$ insertion.
This is because an $LL$ insertion is induced by RG running
from $\Mgrav$ down to the weak scale even in the $RR$ scenario.
The presence of $\ded{23}{LL}$ enhances the effect
of $\ded{23}{RR}$ on \bsbsbar\ mixing,
and $\phi_s$ can be easily pushed to its 95\% probability region.
%% high m0
% Next, we switch to a higher value of $m_0$.
%%% enough contribution to phi_s but likely breaches tmg d_n
% Lower $\tb$ is better.
%%%% LFV close to upper bound
%%%% d_n can be satisfied by modifying d23LL
However, an $RR$ insertion is strictly limited by $d_n$, the neutron EDM\@.
A region allowed by $d_n$ and $\Delta M_s$ around the origin,
is separated from the $\phi_s$ region.
The band obeying $d_n$ can be rotated to overlap the cyan region
by altering $\ded{23}{LL}$ at $\MGUT$, since
$d_n$ is influenced through the combination of
$\mathrm{Im} [\ded{23}{LL} \ded{23}{RR}^*]$.
The presented plots are valid for the phase of $\ded{23}{LL}$ equal to
$\arg (- V_{ts}^* V_{tb})$.
%%% bsg not a big problem
Note that $\bsg$ is not very tight.
This is because
the supersymmetric amplitude does not interfere with the SM one.
%% low tanb
%% high tanb
%%% worsens LFV d_n
LFV and $d_n$ are enhanced for high $\tb$.
Therefore, lowering $\tb$ helps satisfy LFV and $d_n$ as well as $\phi_s$.
One can find that the region preferred by $\phi_s$ involves
the $\tmg$ rate in the vicinity of the current upper limit.
For example, fitting the central value of $\phi_s$ causes
$B(\tmg)$ to be around $10^{-7}$ which is already ruled out by the Belle data.
The area still surviving
could be explored by current and future experiments.
The magnitude of mass insertion accessible with the sensitivity of $10^{-8}$,
attainable at a super $B$ factory, is depicted by a thin circle
inside the current upper bound.
The cyan region is also expected to bring about
$\meg$ at a rate that can be probed by MEG\@.

\begin{figure}
  \centering
  \incgr[height=60mm]{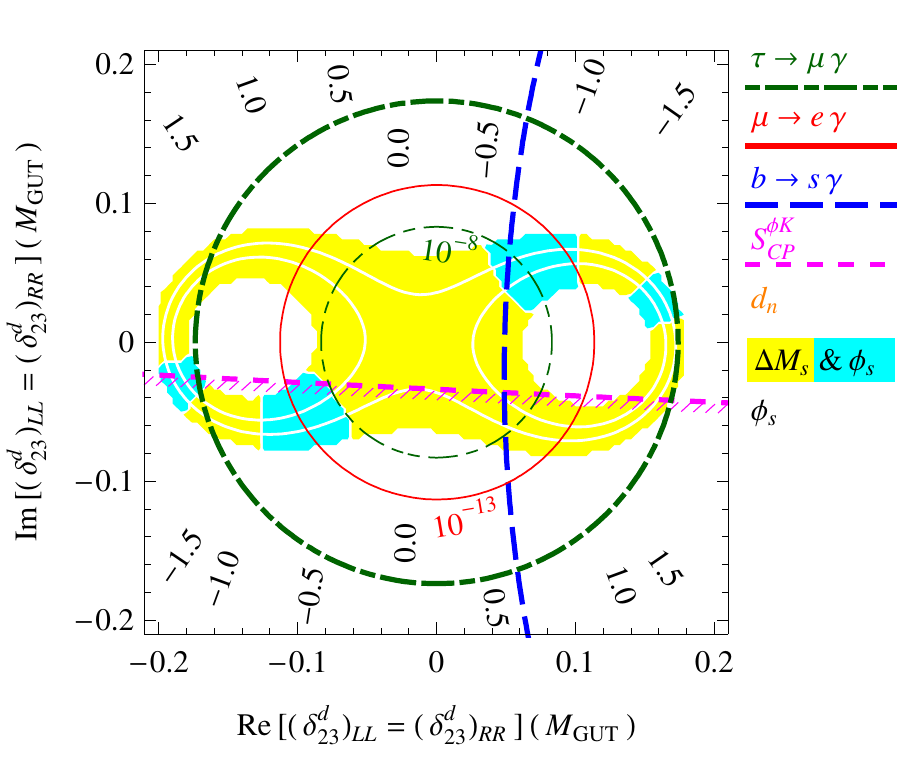}
  \caption{Constraints on $\ded{23}{LL} = \ded{23}{RR}$.
    For $\tmg$, the thick circle is the current upper bound,
    and the thin circle is an upper bound
    from the prospective branching ratio limit, $10^{-8}$.
    For $\meg$, the thin circle shows the projected bound
    on the branching ratio, $10^{-13}$.
    The remaining marks have the same meanings as in Fig.~\ref{fig:RR}.}
  \label{fig:LL=RR}
\end{figure}%
The above restrictions on the $RR$ insertion,
with slight modifications, can be applied
to a popular scenario where
the soft terms are flavor-blind at $\Mgrav$ and
large neutrino Yukawa couplings
are the only source of flavor violation apart from the CKM mixing.
In this case, the slepton mass matrix receives
additional contribution running below $\MGUT$.
Because of this, given the same $\delta$ at $\MGUT$, LFV rates
are higher than in Fig.~\ref{fig:RR}, and therefore one obtains
tighter LFV bounds.

% LL=RR
The last scenario where
$\ded{23}{LL} = \ded{23}{RR}$ at $\MGUT$,
is shown in Fig.~\ref{fig:LL=RR}.
%% low m0
%%% disfavored by phi_s + tmg meg although better than RR
Comparison of Fig.~\ref{fig:LL=RR} and Fig.~\ref{fig:RR} shows
that the conflict between LFV and $\phi_s$
has been reduced here.
%%%% needed mixing smaller than RR
Simultaneous presence of $LL$ and $RR$ mixings
reinforces contribution to the $B_s$ mixing
even with a smaller size of each insertion about 0.1,
while the LFV bounds remain almost the same.
%% low tanb
%%% can fit phi_s LFV d_n
Fig.~\ref{fig:LL=RR} shows regions well inside
the LFV bounds which lead to $\phi_s$ in perfect agreement with
the latest world average.
Part of those regions can satisfy $\SphiK$ and $d_n$ as well.
An area preferred by $\phi_s$ gives rise to $B(\tmg)$ around $10^{-8}$
or larger.
The rate of $\meg$ expected from the same area
is around the sensitivity of MEG\@.
%% high tanb
%%% tanb = 10 still ok
%%%% modifying phase of d23LL helps
% prediction of LFV rates from phi_s
In this scenario, a higher $\tb = 10$ is viable as well \cite{fcncgut}.

% overall observation
%% write in conclusion?

%%%%%%%%%%%%%%
% Conclusion %
%%%%%%%%%%%%%%

We summarize.
% we have studied consequences of the latest $\phi_s$ data on
% LFV in supersymmetric SU(5) GUT, taking a model independent approach.
We have examined three patterns of $\ded{23}{LL}$ and $\ded{23}{RR}$:
$LL$, $RR$, and $LL=RR$.
For reconciling $\phi_s$ with LFV,
it greatly helps to choose the optimal value of
the GUT scale gaugino to scalar mass ratio,
in all these three scenarios.
It appears that the most adequate to fit the current value
of $\phi_s$ is $LL=RR$ among the three scenarios.
The main difficulties
for this purpose are $\bsg$ and $\SphiK$ in the $LL$ scenario,
and LFV and the neutron EDM in the $RR$ scenario.
Inclusion of Planck-suppressed non-renormalizable terms for fixing
the quark--lepton mass relations, in general, affects a LFV bound.
This alteration can be estimated by weakening a $\tmg$ bound
to that from $\tau \rightarrow (e + \mu)\,\gamma$.
In the two scenarios involving an $RR$ mixing,
this reduces the tension between LFV and $\phi_s$.
In all cases, low $\tb$ loosens $\bsg$, $\SphiK$, and $d_n$ as well as LFV,
providing for more room to accommodate $\phi_s$.

%%%%%%%%%%%%%%%%%%%%%%%%%%%%%%%%%%%%%%%%%%%%%%%%
%% BACKMATTER
%%%%%%%%%%%%%%%%%%%%%%%%%%%%%%%%%%%%%%%%%%%%%%%%

\begin{theacknowledgments}
The author thanks P.~Ko and Masahiro Yamaguchi for the collaboration.
He acknowledges Research Grants funded jointly by the Italian
Ministero dell'Istruzione, dell'Universit\`{a} e della Ricerca (MIUR),
by the University of Padova and
by the Istituto Nazionale di Fisica Nucleare (INFN) within the
\textit{Astroparticle Physics Project} and the FA51 INFN Research Project.
This work was supported in part by the European Community Research
Training Network UniverseNet under contract MRTN-CT-2006-035863.
\end{theacknowledgments}

\end{document}